\documentclass[12pt, letterpaper]{article}

\usepackage{amsmath, amsfonts, amscd, amssymb}
\usepackage{amsthm}
\usepackage[all,cmtip]{xy}

\newcommand{\slie}{\mathcal{L}}
\newcommand{\jl}{\mathcal{A}}

\newcommand{\com}{\mathbb{C}}
\newcommand{\A}{\mathbb{A}}
\newcommand{\mapto}{\longrightarrow}

\newcommand{\N}{\mathbb{N}}
\newcommand{\Z}{\mathbb{Z}}
\newcommand{\R}{\mathbb{R}}
\newcommand{\de}{\delta}
\newcommand{\jor}{\mathcal{J}}
\newcommand{\quat}{\mathbb{H}}
\newcommand{\cayl}{\mathbb{O}}
\newcommand{\sed}{\mathbb{S}}
\newcommand{\trig}{\mathbb{T}}

\newcommand{\JL}{\mathfrak{J}\mathfrak{L}}

\title{\bf Multiplicatively Ordered and Directed Hybrid $\mathbf{\de}$-Jordan-Lie
Superalgebra\thanks{In belated loving memory of {\em Professor Joachim (Jim) Lambek}: teacher, university undergraduate academic advisor, research colleague, (homological) algebra mentor, and cordial friend.}}

\author{Ioannis Raptis\thanks{Supply \& Substitute Secondary School Teacher of Mathematics, Physics and Chemistry, Reeson Education, London, United Kingdom; email: {\it irapti11@gmail.com}}}

\date{Latest Version: Thursday 18.7.2024}

\begin{document}

\maketitle

\pagestyle{myheadings}\markboth{\centerline {\small {\sc
{Ioannis Raptis}}}}{\centerline
{\footnotesize {\sc {Ioannis Raptis: Hybrid $\mathbf{\de}$-Jordan-Lie
Superalgebra}}}}

\pagenumbering{arabic}

\begin{abstract}

\noindent A new algebra $\A$, hitherto not encountered in the usual Lie algebraic varieties or supervarieties, is introduced. This paper explores the rich and novel structure of the algebra, and it compares it on the one hand with the Jordan-Lie superalgebras studied by Okubo and Kamiya in \cite{okam}, and on the other, with the four usual Euclidean division rings of the reals ($\R$), the complexes ($\com$), the quaternions ($\quat$) and the octonions ($\cayl$), key algebraic properties of which the algebra is seen to combine, modify, extend and generalise. We especially focus on the novel $\Z_{2}$-graded associative and ordered binary multiplication structure of $\A$ and we compare it to the well known Cayley-Dickson type of extensions of the nonassociative octonions $\cayl$ to the sedenions $\sed$ and, in turn, of the sedenions to the trigintaduonions $\trig$, losing some key algebraic product associativity properties in the process. A potential physical application of the algebra analogous to how the algebra of quaternions has been used in the past to represent the Dirac equation for the free electron \cite{lambek, lambek2} is briefly alluded to at the end.

\end{abstract}

\vskip 0.1in


\hskip 0.1in

\noindent{\footnotesize\underline{\bf New Key words}: Hybrid $\de$-Jordan-Lie Superalgebra; $\Z_{2}$-Graded Associativity; Normally/Lexicographically Ordered and Directed Algebraic Binary Product; Multiplicatively Normally Ordered, $\Z_{2}$-Graded Associative, Lie Admissible, Free Linear Semigroup}

\newpage

\section*{0 \hskip 0.1in Paper R\'{e}sum\'{e}\footnote{The following is an extended version of the {\em Abstract} of the paper as it appears in the Mathematical Physics e-archives {\bf www.arXiv.org}: {\it math-ph: 2405.01181} (v2).}}

\vskip 0.1in

In the present paper, a new set of algebraic numbers $\A$ is introduced. $\A$ is a four-dimensional algebra hitherto 
not encountered in either the usual
algebraic varieties or supervarieties. The algebra is a
$\Z_{2}$-graded and multiplicatively deformed version of the
quaternions $\quat$, with structure similar to that of a $\de$-Jordan-Lie
algebra as defined in \cite{okam}, but it is shown to be neither
that of a purely associative ($\de=+1$) Lie superalgebra, nor that
of a purely antiassociative ($\de=-1$) Jordan-Lie superalgebra.
Rather, it exhibits a novel kind of associativity, here called
{\em multiplicatively normally ordered $\Z_{2}$-graded associativity}, that is somewhat
a hybrid between pure associativity and pure antiassociativity. 

In addition to $\Z_{2}$-graded associativity, the generators of $\A$ obey
graded commutation relations encountered in both the usual
$\Z_{2}$-graded Lie superalgebras ($\de=1$) and in $\Z_{2}$-graded
Jordan-Lie superalgebras ($\de=-1$). They also satisfy a novel type of {\em ordered $\Z_{2}$-graded 
Jacobi identities} that combine characteristics of the Jacobi
relations obeyed by the generators of ungraded Lie,
$\Z_{2}$-graded Lie and $\Z_{2}$-graded Jordan-Lie algebras.
Mainly due to these three structural algebraic features, $\A$ is called a {\em hybrid
$\de$-Jordan-Lie superalgebra}. 

Additionally, we present a heuristic and intuitive argument of how $\A$ may arise in a similar way to how the real numbers $\R$ extend to the complexes $\com$; the complexes $\com$ to the quaternions $\quat$, and the quaternions $\quat$ to Cayley's octonions $\cayl$, with every time each extension, which doubles the dimensionality of the corresponding vector space, being accompanied by a loss of some important (algebraic) structure. In fact, $\A$ is seen to combine quintessential algebraic characteristics of all the four existing Euclidean division rings \cite{hur}: $\R$, $\com$, $\quat$ and $\cayl$, thus further corroborating that it is a $4$-fold hybrid of them all, apart from its additional super Lie algebra-like characteristics. 

We also witness that $\A$, unlike the division algebras $\com$, $\quat$ and $\cayl$ that it combines and extends, is non-involutive ({\it i.e.}, it is not a $\star$-algebra), hence it derives its metric-norm directly from its binary product alone and not from a $\star$-involution ({\it i.e.}, the $\com$-conjugation unary operation observed in $\com$, $\quat$ and $\cayl$); moreover, its metric is non-Euclidean: it is a traceless Kleinian pseudo-metric $\eta_{\mu\nu}=(1,-1, 1, -1)$, of signature $0$.

Also, as a result of its lack of $\star$-involution and a glaring absence of a two-sided (left/right) identity generator, $\A$ is seen not to have formal multiplicative inverses of its generators, thus it is not a division ring. By contrast, the absence of inverses renders $\A$ to a novel algebraic structure, here coined {\em one-sided identity, free generative, non-associative linear semigroup}, in which product strings of $\A$'s generators are viewed as words, which can then linearly combine over $\R$ to form linear superpositions thereof. The said multiplication order defines $\A$ as a novel kind of {\em alphabetic algebra}, hereby coined {\em the alphabet algebra} $\A$, whose structure as a $\Z_{2}$-graded associative and as a $\Z_{2}$-graded Jordan-Lie superalgebra vitally depends on that lexicographic order of multiplication of the letter generators in its algebraic words. 

The present paper defines $\A$, works out its structure as a {\em hybrid
$\de$-Jordan-Lie superalgebra} following Okubo and Kamiya in \cite{okam}, it then 
compares it with the $\de$-Jordan-Lie superalgebra defined there, and it further abstracts and generalises it to a new category of {\em multiplicatively ordered (lexicographic) graded Lie algebraic-cum-nonassociative linear semigroup} supervarieties. 

Due to its novel multiplicatively ordered $\Z_{2}$-graded associative structure, we especially dwell on the (non)associativity properties of a general algebraic binary product, such as {\em alternativity, power associativity, flexibility, composability relative to a norm and Lie algebra admissibility}, and we compare $\A$ to the four Euclidean division algebras ($\R$, $\com$, $\quat$, $\cayl$), their further extensions to the {\em sedenions} $\sed$ \cite{imaeda, sed} and the {\em trigintaduonions} $\trig$ \cite{trig1, trig}, as well as to the general {\em nonassociative Lie-type Okubo algebras} \cite{okubo1,okubo2,okubo3}. 

The algebra $\A$ is of mathematical interest in its own right and the present paper mainly explores the new mathematical import of $\A$; however, at the end of the paper we discuss a couple of potential physical applications that have been intuited ever since its original inception by the present author three decades ago in the course of writing his Ph.D. thesis \cite{rap1}. We leave more detailed elaborations on the physical import and applications of $\A$ to a forthcoming paper \cite{rap2}.

\section{Paper Overview cum Introductory Remarks on Lie Superalgebras}\label{sec1}

In this opening section, we first recall the structure of $\Z_{2}$-graded Lie algebras, commonly known as {\em Lie superalgebras} \cite{freund}. 

In Section 2, we recall the definition of an abstract
$\de$-Jordan-Lie ($\de$-J-L) algebra $\jl$ as given by Okubo and
Kamiya in \cite{okam}, which, as we shall see, includes as a
particular case the $\Z_{2}$-graded Lie superalgebra to be defined below. 

We then introduce the concrete Hybrid Jordan-Lie superalgebra $\A$
(Section \ref{sec3}), and finally we compare the key defining
properties of the two structures (Section \ref{sec4}). 

Here, in Section \ref{sec4}, we abstract and generalise $\A$ on two fronts:

\begin{itemize}

\item As {\em an abstract Hybrid $\Z_{2}$-graded associative Jordan-Lie Superalgebra};

\item As {\em an abstract multiplicatively ordered and directed, $\Z_{2}$-graded associative free linear semigroup}.

\end{itemize}

Section \ref{sec5} compares the structure of $\A$ against the four usual Euclidean division algebras: the reals ($\R$), 
the complexes ($\com$), the quaternions ($\quat$) and the octonions ($\cayl$). We give an informal and heuristic Cayley-Dickson type of ring extension 
procedure of how $\R$ can be extended, via $\com$ and $\quat$, all the way to $\cayl$ by losing some important algebraic structure each step along the way. In particular, we dwell on the nonassociative character of $\A$ and how this compares to the nonassociativity of the octonions $\cayl$ \cite{conway, lounesto, springer}, and beyond: to the algebras of {\em sedenions} $\sed$ \cite{imaeda, sed} and {\em trigintaduonions} $\trig$ \cite{trig1, trig}, which are further Cayley-Dickson type of extensions-complexifications of the $8$-dimensional $\cayl$ to $16$ ($\sed$) and $32$ ($\trig$) dimensions, respectively. Furthermore, by $\A$'s {\em Lie admissibility}, its non-associativity, its not having a 2-sided identity element, its being {\em non-involutive} and its supporting a non-Euclidean type of metric-norm, we liken it to {\em Okubo algbebras} \cite{okubo1,okubo2,okubo3}, albeit, {\em multiplicatively ordered} and $\Z_{2}$-graded associative ones, as defined in the present paper. 

We conclude the paper (Section 6) with some brief remarks about a
possible physical application and interpretation of $\A$, the details and full elaboration of which we leave for a forthcoming paper \cite{rap2}.

\subsection{$\Z_{2}$-graded Lie Superalgebras}

In theoretical physics, supersymmetry pertains to a symmetry
between bosons and fermions \cite{freund}. Supergroups, or $\Z_{2}$-graded Lie
groups, are the mathematical structures modelling continuous
supersymmetry transformations between bosons and fermions. As Lie
algebras consist of generators of Lie groups---the infinitesimal
Lie group elements tangent to the identity \cite{lie1}, so $\Z_{2}$-graded Lie
algebras, otherwise known as Lie superalgebras, consist of
generators of (or infinitesimal) supersymmetry transformations
\cite{freund}.

Like their ungraded Lie ancestors $L$, Lie superalgebras $\slie$

\begin{itemize}

\item (i) Are real or complex vector spaces that are $\Z_{2}$-graded\footnote{It is tacitly assumed
that both $\slie^{0}$ and $\slie^{1}$ in (\ref{eq1}) are linear
subspaces of $\slie$ whose only common element is the zero vector
$0$. $\slie^{0}$ is usually referred to as {\em the even subspace
of} $\slie$, while $\slie^{1}$ as {\em the odd subspace of}
$\slie$.}

\begin{equation}\label{eq1}
\slie=\slie^{0}\oplus\slie^{1},
\end{equation}

\noindent with grading function $\pi$ given by

\begin{equation}\label{eq2}
\pi(x):=\left\lbrace\begin{array}{rcl} 0,& \mbox{when}~
x\in\slie^{0},\cr 1, & \mbox{when}~ x\in\slie^{1}.
\end{array} \right.
\end{equation}

\item (ii) Are associative algebras with respect to a bilinear product
$\cdot:~\slie\otimes\slie\mapto\slie$ (simply write $x\cdot
y\equiv xy=z\in\slie$ for the associative product $\cdot$ of $x$
and $y$ in $\slie$).

\item (iii) Close under the so-called super-Lie bracket $<.
,.>:~\slie\otimes\slie\mapto\slie $ represented by the
non-associative, bilinear, $\Z_{2}$-graded (anti-)commutator Lie
product $[.,.\}$ defined as

\begin{equation}\label{eq3}
[x,y\}:=\left\lbrace\begin{array}{rcl} [x,y]=xy-yx\in\slie^{0},&
\mbox{when}~ x,y\in\slie^{0},\cr \{ x,y\}=xy+yx\in\slie^{0}, &
\mbox{when}~ x,y\in\slie^{1},\cr [x,y]=xy-yx\in\slie^{1}, &
\mbox{when}~ x\in\slie^{0}~{\rm and}~ y\in\slie^{1}.
\end{array} \right.
\end{equation}

\item (iv) With respect to $<.,.>$, they obey the so-called
super-Jacobi identities.\footnote{For more details about the
properties (i)--(iv) of Lie superalgebras, the reader is referred
to \cite{freund}. We will encounter them in a slightly different
guise and in more detail when we define $\de$-Jordan-Lie
superalgebras in the next section.}

\end{itemize}

We now turn our attention on the $\de$-Jordan-Lie Superalgebras studied in \cite{okam}.

\section{$\de$-Jordan-Lie Superalgebra}\label{sec2}

Let $\jl$ be a finite dimensional vector space over a field $K$ of
characteristic not $2$ which, for familiarity, one may wish to
identify with $\R$ or $\com$. Also, let $\jl$ be $\Z_{2}$-graded

\begin{equation}\label{eq4}
\jl=\jl^{0}\oplus\jl^{1},
\end{equation}

\noindent with grader $\pi$ given by

\begin{equation}\label{eq5}
\pi(x):=\left\lbrace\begin{array}{rcl} 0,& \mbox{when}~
x\in\jl^{0},\cr 1, & \mbox{when}~ x\in\jl^{1},
\end{array} \right.
\end{equation}

\noindent as in (\ref{eq1}) and (\ref{eq2}) for $\slie$
above.\footnote{In \cite{okam}, $\sigma(x)$ is used instead of
$\pi(x)$ to symbolise the grading function. See $(1.2)$ in
\cite{okam}. }

Next, we consider only homogeneous elements of $\jl$ ({\it i.e.},
either $x\in\jl^{0}$ or $x\in\jl^{1}$, but not $z=\alpha x+\beta
y,~ x\in\jl^{0},~ y\in\jl^{1};~\alpha ,\beta\in \com$),\footnote{In
theoretical physics, this forbidding of linear combinations
between bosons and fermions is known as the Wick-Wightman-Wigner
superselection rule \cite{www}. The direct sum split between the
even and the odd subspaces in (\ref{eq1}) and (\ref{eq4}) is
supposed to depict precisely this constraint to free
superpositions between quanta of integer and half-integer spin
({\it i.e.}, bosons and fermions, respectively). Mainly because of
\cite{www} we decided to symbolise the grading function in
(\ref{eq2}) and (\ref{eq5}) by `$\pi$' (for `{\em intrinsic
parity}') rather than by `$\sigma$' as in \cite{okam}. In the
literature, the set-theoretic (disjoint) union `$\cup$' is
sometimes used instead of `$\oplus$' between the even and odd
subspaces of a $\Z_{2}$-graded vector space \cite{freund}---it
being understood that these two subspaces have only the trivial
zero ($0$) vector in common, as noted in footnote 1. `$\cup$' too is
supposed to represent the aforesaid spin-statistics superselection
rule.} and as in $(1.3)$ of \cite{okam} we define

\[
(-1)^{xy}:=(-1)^{\pi(x)\pi(y)}.
\]

Let also $xy$ be a bilinear product in $\jl$ satisfying

\begin{equation}\label{eq6}
(xy)z=\de x(yz),~\de=\pm1,
\end{equation}

\noindent with respect to which $\jl$ is said to be a {\em
$\de$-associative algebra}. In particular, for $\de=+1$, $\jl$ is
an {\em associative} algebra; while for $\de=-1$, it is {\em
antiassociative}.

Consider also a second bilinear product $<.,.>:\, \jl\otimes
\jl\mapto\jl$

\begin{equation}\label{eq7}
<x,y>:=xy-\de(-1)^{xy}yx,
\end{equation}

\noindent satisfying

\begin{equation}\label{eq8}
\pi(<x,y>)=\pi(x)+\pi(y)~~({\rm mod}~ 2),
\end{equation}

\begin{equation}\label{eq9}
<x,y>=-\de(-1)^{xy}<y,x>,
\end{equation}

\noindent and

\begin{equation}\label{eq10}
(-1)^{xz}<<x,y>,z>+(-1)^{yx}<<y,z>,x>+(-1)^{zy}<<z,x>,y>=0,
\end{equation}

\noindent or equivalently

\begin{equation}\label{eq11}
(-1)^{xz}<x,<y,z>>+(-1)^{yx}<y,<z,x>>+(-1)^{zy}<z,<x,y>>=0.
\end{equation}

$\jl$, satisfying (\ref{eq4})--(\ref{eq11}), is called a {\em
$\de$-J-L algebra} \cite{okam}. Also, one can easily verify that
for $\de=1$, $\jl$ is the associative $\Z_{2}$-graded Lie
superalgebra $\slie$ defined in (i)--(iv) of section
\ref{sec1}.\footnote{In particular, the expression (\ref{eq3}) in
(iii) is encoded in (\ref{eq7})--(\ref{eq9}) above, while the
`graded Jacobi identities' property (iv) of $\slie$ is expressed
by (\ref{eq10}) or (\ref{eq11}).} The antiassociative ($\de=-1$)
case is coined {\em Jordan-Lie superalgebra} in \cite{okam}---here
to be referred to as {\em J-L algebra} $\jor$ for short. We may
summarise all this as follows

\[
\jl=\left\lbrace\begin{array}{rcl} \slie, & \mbox{for}~ \de=+1,\cr
\jor, & \mbox{for}~ \de=-1.
\end{array} \right.
\]

For future use, we quote, without proof, the following lemma and
two corollaries from \cite{okam}:\footnote{Proofs can be read
directly from \cite{okam}.}

\begin{itemize}

\item {\bf Lemma:} {\em In every antiassociative algebra $A$, any product involving
four or more elements of $A$ is identically zero}.\footnote{Lemma
$1.1$ in \cite{okam}.}

\item {\bf Corollary 1:} {\em Antiassociative algebras have no
idempotent elements and, as a result, no units ({\it i.e.}, identity
elements)}.\footnote{Corollary $1.2$ in \cite{okam}.}

\item {\bf Corollary 2:} {\em Let $\jor$ be a J-L algebra as defined
above. Then $\jor$ is nilpotent of length at most 3} (write:
$\jor_{4}=0$).\footnote{Corollary $1.1$ in \cite{okam}.}

\end{itemize}

\section{Introducing the Alphabetic Hybrid Jordan-Lie Superalgebra $\A$}\label{sec3}

Let $\A$ be a $4$-dimensional vector space over $\R$ spanned by
$\mathcal{G}=\{ a,b,c,d\}$\footnote{The alphabetic symbolism of
the four basis vectors (generators) in $\mathcal{G}$ will be
explained subsequently.} and also be $2\oplus2$-dimensionally
$\Z_{2}$-graded thus

\begin{equation}\label{eq12}
\A=\A^{0}\oplus\A^{1}=\mathrm{span}_{\R}\{ a
,b\}\oplus\mathrm{span}_{\R}\{ c ,d\} .
\end{equation}

Let $\circ:~\A\otimes\A\mapto\A$ be a bilinear product that closes in $\A$
which, in terms of $\A$'s generators in $\mathcal{G}$, is encoded
in the following $(4\times 4)$-multiplication (:product) table

\begin{equation}\label{eq13}
\begin{tabular}{|c||c|c|c|c|}
\hline
$\circ$ &$a$ &$b$ &$c$ &$d$ \\ \hline\hline $a$ &$a$ &$b$ &$-d$ &$-c$ \\
\hline $b$ &$b$ &$-a$ &$-d$ &$c$ \\ \hline $c$ &$c$
&$d$ &$a$ &$-b$ \\ \hline $d$ &$d$ &$-c$ &$b$ &$-a$ \\
\hline
\end{tabular}
\end{equation}

\vskip 0.05in

From table (\ref{eq13}), one can straightforwardly extract the following
information:

\begin{itemize}

\item {\em The binary product $\circ$ is not commutative}. In particular, $a$ commutes only with
$b$; while, $b$, $c$ and $d$ mutually anticommute. Moreover, $a$
is a right-identity, but not a left one.

\item {\em $\circ$ is not (anti)associative}. For example, one can evaluate

\[
c=-ad=a(bc)\not=\left\lbrace\begin{array}{rcl} (ab)c=bc=-d, &
(\de=+1);\cr -(ab)c=-bc=d, & (\de=-1).
\end{array} \right.
\]

\item $a$ and $c$ are $\sqrt{a}$, while $b$ and $d$ are
$\sqrt{-a}$.

\item The even subspace of $\A$ in (\ref{eq12}), $\A^{0}:=\mathrm{span}_{\R}\{
a,b\}$, is isomorphic to the complex numbers $\com$ if one makes the
following correspondence between the unit vectors (generators) of $\A$ and
$\com$

\[
\A^{0}\ni a\mapto 1\in\com~\mathrm{and}~\A^{0}\ni b\mapto i\in\com~ (i^{2}=-1~\mathrm{and}~b^{2}=-a).
\]

\noindent with $\A^{0}$ being the subalgebra of even elements of $\A$.

\item The product of an even and an odd generator is odd, while
the product of two odd generators is even. Together with the
second observation above, we may summarise this to the following

\[
\pi(xy)=\pi(x)+\pi(y)~~({\rm mod 2}).
\]

\item The inhomogeneous vector $\mathbf{n_{1}}=b+c$ and the odd vector
$\mathbf{n_{2}}=c+d$ are nilpotent.\footnote{We will return to these two vectors in the last section when we discuss norm issues in $\A$. 
The reader should note that $\mathbf{n_{1}}$
violates the aforementioned Wick-Wightman-Wigner spin-statistics superselection
rule \cite{www} as it linearly combines vectors in the even and the odd subspaces of $\A$.}

\end{itemize}

\subsection{`Naive' Cayley-Dickson type division ring extension heuristics}

Let us try to gain some more insight into the non-associativity of $\circ$ by
making a formal correspondence between the `units' of $\A$ in
$\mathcal{G}$ and the standard unit quaternions $\mathcal{U}=\{ 1,
i, j, k\}$ in $\quat$

\begin{equation}\label{eq14}
\begin{array}{c}
a\mapto 1,~~ b\mapto i,\cr c\mapto j,~~d\mapto k.
\end{array}
\end{equation}

\noindent Then, one may wish to recall that the {\em associative
division algebra} $\quat$\footnote{We may write $\bullet$ for the
associative binary product of quaternions ({\it i.e.},
$\bullet:~\quat\otimes\quat\mapto \quat$), but omit it in actual
products, that is to say, we simply write $xy$ ($x,y\in\quat$). We
assumed the same thing for $x\cdot y$ in $\slie$ and $\jl$, as
well as for $x\circ y$ in $\A$ (for instance, see (ii) in section
\ref{sec1}).} can be obtained from $\com$ by adjoining
$j=\sqrt{-1}$ to the generators $\{ 1, i\}$ of $\com$ and by
assuming that it commutes with $1$ 

$$1j=j1=j$$

\noindent but that {\em it anticommutes
with} $i$ and that {\em it closes in} $\quat$\footnote{That is to say, $ij=k$ is also a unit quaternion generator in $\quat$, thus completing the set of four unit quaternion generators $\{ 1, i, j, k\}$.}

$$ij=-ji=k\in\quat$$

\noindent In fact, one assumes that by
transposing $i$ with $j$, $i$ gets conjugated \cite{kauff, conway, springer}, as follows

$$ij=ji^{\star}=-ji\Leftrightarrow\{ i,j\}:=ij+ji=0$$

\noindent Then, {\em by assuming associativity, one verifies that $k$ too is a}
$\sqrt{-1}$, {\em that also anticommutes with both} $i$ and $j$

\[
\begin{array}{c}
k^{2}=(ij)(ij)=i(ji)j=-i^{2}j^{2}=-1,\cr
ki=(ij)i=i(ji)=-i(ij)=-ik\Leftrightarrow\{ i,k\}:=ik+ki=0
\end{array}
\]

\noindent thus one completes the following well-known
multiplication table for the four unit quaternions 

\begin{equation}\label{eq15}
\begin{tabular}{|c||c|c|c|c|}
\hline
$\bullet$ &$1$ &$i$ &$j$ &$k$ \\ \hline\hline $1$ &$1$ &$i$ &$j$ &$k$ \\
\hline $i$ &$i$ &$-1$ &$k$ &$-j$ \\ \hline $j$ &$j$
&$-k$ &$-1$ &$i$ \\ \hline $k$ &$k$ &$j$ &$-i$ &$-1$ \\
\hline
\end{tabular}
\end{equation}

$\bullet$ {\bf A passing note on `3-chirality'.} At this point, it is important to note that in the case of the quaternions $\quat$, from (\ref{eq15}) one assumes the `left chiral' order of multiplication for the anti-commuting unit quaternions: $ijk=-1$;\footnote{The epithet `left/right' chiral for the multiplcation orders $ijk=-1$ and its opposite $kji=1$ respectively arises from the fact that the three mutually anticommuting unit quaternions (or equivalently, the Pauli spin matrices) geometrically represent rotations in $\R^{3}$, which can be set to be clockwise or anti-clockwise \cite{lambek, conway}.} while, in the case of our $\A$, and in view of the formal correspondence between the respective $\quat$ and $\A$ generators that we made in (\ref{eq14}), one can see straight from the binary multiplication table of $\A$ in (\ref{eq13}) that we assume the opposite `right chiral' order of multiplication of the three anticommuting unit generators of $\A$: $dcb=-a$. The `reason' for assuming this multiplication order and not its opposite $bcd=a$ will become transparent as the paper unfolds below. 

$\bullet$ {\bf The crux of the argument.} If we were to emulate the naive and heuristic Cayley-Dickson extension of $\com$ to $\quat$ shown above in the
case of $\A$, thus adjoin $c$ to $b$ in $\A^{0}\simeq\com$ and
require according to (\ref{eq13}) that they anticommute, as well
as that {\em the binary product $\circ$ be associative}, we would get

\begin{equation}\label{eq16}
d^{2}=(cb)(cb)=cbcb=-c^{2}b^{2}=-(a)(-a)=a
\end{equation}

\noindent which disagrees with entry $(4,4)$ in table
(\ref{eq13}). Similarly for the generator $c$.\footnote{We encourage the reader to try to calculate
$c^{2}=(bd)(bd)$ in a manner similar to (\ref{eq16}) above.} 
Clearly then, as also noted above, (the product $\circ$ in) $\A$
is neither associative\footnote{$\de=1$ in (\ref{eq6}).} nor
antiassociative\footnote{$\de=-1$ in (\ref{eq6}).}

\begin{quotation}

\noindent $\bullet$  \underline{\bf Question:} {\em How can we obtain agreement between
products like the one in} (\ref{eq16})---which arise rather
naturally upon trying to extend $\com$ to $\A$ in the same manner
that $\com$ is extended to $\quat$ \cite{kauff, conway, springer}---{\em with the entries of the
multiplication table} (\ref{eq13})? Evidently, {\em we need a new
(anti)associativity-type of law for the binary product} $\circ$.

\end{quotation}

\subsection{$\A$ viewed as a multiplicatively ordered, freely generated algebra}

\noindent To the end of the question posed above, one might first wish to regard $\A$ as a {\em free algebraic structure},\footnote{Broadly speaking, a free algebraic structure $\mathcal{A}$ is a set of generating elements $\mathcal{G}$ called {\em letters}, endowed with a binary multiplication operation---the algebraic product concatenation of the letter-generators in $\mathcal{G}$---that is subject to certain `constraints', formally called {\em rules} or {\em relations}, for the formation and evaluation (:contraction) of product strings of letters called {\em words}. See next.} and first define:

\vskip 0.1in

{\bf Definition 1:} A product string $w$ of generators of $\A$ in
$\mathcal{G}$ of length $l$ greater than or equal to
$3$\footnote{As noted in the previous footnote, in {\em free algebra} jargon, such product strings
$w$ are called {\em words} and their factors, which are elements
of $\mathcal{G}$, are called {\em letters} (which, in turn, makes
$\mathcal{G}$ $\A$'s $4$-letter alphabet!). The number $l$ of
letters in a word $w$ is its {\em length}, and we write $l(w)$.
Formally speaking, a word $w$ of length $l$ is a member of
$\stackrel{l~{\rm factors}~\A}
{\overbrace{\A\otimes\A\otimes\cdots\otimes\A}}$. The $4^{2}=16$
possible words of length $2$ in $\A$ are the ones depicted in the multiplication 
table (\ref{eq13}) above.} is said to be {\em ({\sl N})ormal ({\sl T})ime
({\sl O})rdered}\footnote{Write `{\sl NTO}-ed' and symbolise the
word by $\overleftarrow{w}$. We originally encountered this term in this author's doctoral thesis \cite{rap1} where we borrowed it from Quantum Field Theory jargon \cite{haag, str}, as a conflation of the usual {\sl normally} and {\sl time} ordered products of quantum field operators there.} if it is of the following `{\em
right-to-left alphabetical order}' or `{\em lexicographic-syntax}'

\begin{equation}\label{eq17}
\overleftarrow{w}:=d^{s}c^{r}b^{q}a^{p},~~p,q,r,s\in\N;~~l(w):=p+q+r+s.
\end{equation}

\noindent Then we impose the following three {\em rules} or {\em
relations}\footnote{Again, this is free algebra jargon.} onto the
total contraction of any word of length $l\geq3$:\footnote{By `{\em
total contraction}' of a word of length $l\geq3$ we mean the
evaluation-reduction of the word to a single signed ($\pm$) letter in $\mathcal{G}$
after $l-1$ pairwise contractions of its constituent letters
according to (\ref{eq13}). Again, formally speaking, the product
$\circ:~\A\otimes\A\mapto\A$ in (\ref{eq13}) represents the
contraction of $2$-words in $\A$, so analogously, the total
contraction of words of length $l$ may be cast as $\circ^{l-1}:~
\stackrel{l-1~{\rm times}}
{\overbrace{\A\otimes\A\otimes\cdots\otimes\A}}\mapto\A$.}

\vskip 0.1in

{\bf Rule 0:} Before contracting totally a word $w$ of length
$l\geq3$, it should be brought into {\sl NTO}-ed form in the
following two steps:

\begin{itemize}

\item (a) When the right-identity letter $a$ is found in an extreme
left or intermediate position in $w$, it should be contracted with
the adjacent letter on its right according to
(\ref{eq13}).\footnote{As it were, the `natural' position of $a$ in
a word is to the extreme right. This seems to suit $a$'s role as a
right-identity in $\A$ (\ref{eq13}).}

\item (b) The other three mutually anticommuting generators $b$, $c$ and $d$
in $\mathcal{G}$ should be pairwise swapped within $w$ so that they
are ultimately brought to the form $\pm d^{s}c^{r}b^{q}$.

\end{itemize}

\noindent A couple of comments are due here:

\vskip 0.1in

1) Above, (a) implies that the length of a word may change upon
{\sl NTO}-ing it. This is allowed to happen in $\A$. For the
algebraic structure of $\A$ that we wish to explore here not all
words assembled by free (arbitrary) $\circ$-concatenations of
letters in $\mathcal{G}$ are significant. {\em Only {\sl NTO}-ed
words are structurally significant},\footnote{This will be amply
justified in the sequel.} and any given $w$ has a unique {\sl
NTO}-ed form $\overleftarrow{w}$ fixed according to (i) and (ii)
above. Rule 0 prompts us to call $\A$ `{\em multiplicatively
ordered}' and this alphabetico-syntactic ordering may be formally
cast as follows

\begin{equation}\label{eq18}
\mathrm{Lexicographic\,product\, ordering}\, from-right-to-left: \, d>c>b>a
\end{equation}

\noindent since, once again, every {\sl NTO}-ed word is of the form
$\overleftarrow{w}:=d^{s}c^{r}b^{q}a^{p}$ according to
(\ref{eq17}). The generators of $\A$ are ordered thus. 

It follows that, since in
its transition to its unique {\sl NTO}-ed form a word may change
length, the latter is not a significant structural trait of $\A$,
but the order (\ref{eq18}) is.

2) Normal ordering respects superpositions of words in $\A$. In
other words, {\sl NTO}-ing is a linear operation; symbolically

\[
\overleftarrow{\alpha w_{1}+\beta w_{2}}=\alpha\overleftarrow{w_{1}}+\beta \overleftarrow{w_{2}},\; \alpha,\beta\in\R
\]

\noindent The other two rules that we impose on the total
contraction of a {\em NTO}-ed word of length $l\geq3$ are:

\vskip 0.1in

{\bf Rule 1:} Every {\sl NTO}-ed word of length $l$ greater than 2
contracts fully to a (signed) letter in $\mathcal{G}$ by $l-1$
sequential pair-contractions of letters in it according to
(\ref{eq13}) {\em (f)rom (r)ight (t)o (l)eft}\footnote{Write
`frtl'.} ({\it i.e.}, in the multiplicative order depicted in
(\ref{eq18})). We may call this rule for $\circ$ {\em ordered} or
{\em directed associativity}.

\vskip 0.1in

{\bf Rule 2:} Moreover, ordered associativity is $\Z_{2}$-graded
as follows

\small{
\begin{equation}\label{eq19}
\begin{array}{l}
\overleftarrow{w_{1}}=\ldots oe^{'}e\mapto(\overleftarrow{w_{1}})=
(-1)^{[\pi(e)+\pi(e^{'})]}\ldots o(e^{'}e)=+\ldots oe^{''},~
e^{''}=(e^{'}e)~{\rm from}~(13)\cr \overleftarrow{w_{2}}=\ldots
o^{''}o^{'}o\mapto(\overleftarrow{w_{2}})=(-1)^{[\pi(o)+\pi(o^{'})]}\ldots
o^{''}(o^{'}o)=+\ldots o^{''}e,~e=(o^{'}o)~{\rm from}~(13)\cr
\overleftarrow{w_{3}}=\ldots
o^{'}oe\mapto(\overleftarrow{w_{3}})=(-1)^{[\pi(e)+\pi(o)]}\ldots
o^{'}(oe)=-\ldots o^{'}o^{''}, o^{''}=(oe)~{\rm from}~(13),
\end{array}
\end{equation}}

\noindent where $(\overleftarrow{w})$ signifies the commencement
of the pairwise sequential total contraction of the {\sl no}-ed
word $w$ frtl {\it \`a la} rule 1; `$e$' stands for ($e$)ven and
`$o$' for ($o$)dd letters in $\overleftarrow{w}$; and
`$e^{''}=(e^{'}e)~{\rm from}~(13)$' at end of the first row of
(\ref{eq19}) signifies the contraction and substitution of the
product pair $e^{'}e$ by $e^{''}$ according to
(\ref{eq13}).\footnote{And from now on, $(xy)$ in $\A$ will
indicate precisely this `contraction of $xy$ and its substitution
by the corresponding entry from (\ref{eq13})' process.}. Thus,
rule 2 essentially says that {\em when an odd and an even letter
contract within a {\sl no}-ed word $\overleftarrow{w}$, one must
put a minus sign in front of $\overleftarrow{w}$} In view of
rules 0--2, we call $\circ$ in $\A$ a `{\em $\Z_{2}$-graded
ordered associative product}'. The $\Z_{2}$-graded ordered
associativity of $\A$ is somewhat `in between' the pure
associativity of a Lie superalgebra $\slie$ ($\de=1$) and the pure
antiassociativity of a J-L algebra $\jor$ ($\de=-1$) as defined
above.

\begin{quotation}

\noindent Due to rules 0--2, $\A$ may be called a {\em
multiplicatively ordered $\Z_{2}$-graded associative
algebra}.\footnote{From now on we will most often drop the adverb
`multiplicatively' above and simply refer to $\A$ as an ordered
$\Z_{2}$-graded associative algebra.}

\end{quotation}

Having rules 0--2 in hand, we are now in a position to show that
words such as the one displayed in (\ref{eq16}) contract
consistently with the binary multiplication table (\ref{eq13}),
thus we provide an answer to the question following (\ref{eq16})
above. So, we check that

\begin{equation}\label{eq20}
d^{2}=(cb)(cb)=cbcb\stackrel{R0}{=}-c^{2}b^{2}\stackrel{R1}{=}-c^{2}(b^{2})
\stackrel{(\ref{eq13})}
{=}cca\stackrel{R1}{=}c(ca)\stackrel{R2}{=}-c^{2}\stackrel{(\ref{eq13})}{=}-a,
\end{equation}

\noindent is in agreement with (\ref{eq13}).\footnote{We note that
in (\ref{eq20}), $R0$, for instance, refers to `Rule 0' (similarly
for $R1$ and $R2$). Again, for `practice' the reader can also
verify that $c^{2}=(bd)(bd)=\cdots=a$, in agreement with
(\ref{eq13}).}

\subsection{The Lie admissibility of $\A$}

Now we can give the rest of the $\Z_{2}$-graded Lie algebra-like
structural properties of $\A$.

\begin{itemize}

\item First, there is a bilinear product
$<.,.>:~\A\otimes\A\mapto\A$ represented by the non-associative,
$\Z_{2}$-graded (anti-commutator) Lie product $[.,.\}$ as follows

\begin{equation}\label{eq21}
[x,y\}:=\left\lbrace\begin{array}{rcl} [x,y]=xy-yx\in\A^{0},&
\mbox{when}~ x,y\in\A^{0},\cr \{ x,y\}=xy+yx\in\A^{0}, &
\mbox{when}~ x,y\in\A^{1},\cr \{ x,y\}=xy+yx\in\A^{1}, &
\mbox{when}~ x\in\A^{0}~{\rm and}~ y\in\A^{1},
\end{array} \right.
\end{equation}

\noindent which is similar to (\ref{eq3}), and it also satisfies 

\begin{equation}\label{eq22}
\pi(<x,y>)=\pi(x)+\pi(y)~~({\rm mod}~2)
\end{equation}

\noindent as well as

\begin{equation}\label{eq23}
\begin{array}{c}
<x,y>:=xy-\de(-1)^{xy}yx=-\de(-1)^{xy}<y,x>=\cr=\left\lbrace\begin{array}{rcl}
xy-(-1)^{xy}yx,& \mbox{when}~ x,y\in\A^{0}~{\rm
or}~x,y\in\A^{1};~~\underline{\de=1},\cr xy+(-1)^{xy}yx, &
\mbox{when}~ x\in\A^{0}~{\rm and}~
y\in\A^{1};~~\underline{\de=-1},
\end{array} \right.
\end{array}
\end{equation}

\noindent similar to (\ref{eq7}), (\ref{eq8}) and
(\ref{eq9}), respectively.\footnote{We will comment further on (\ref{eq21}) and
(\ref{eq22})-(\ref{eq23}) in the next section when we compare $\A$
and the $\de$-J-L algebra $\jl$ of \cite{okam}.}

\item Second, the following eight possible super-Jacobi identities

\begin{equation}\label{eq24}
\begin{array}{rcl}
&[\{ d ,c\} ,a]+\{\{ c ,a\} ,d\}+\{\{ a,d\} ,c\}=0,\cr &[\{d,c\}
,b ]+\{\{ c,b\} ,d\}+\{\{ b,d\} ,c\}=0,\cr &\{ [a,b],d\}+\{\{
b,d\} ,a\}+\{\{ d,a\} ,b\}=0,\cr &\{ [a,b],c\}+\{\{ b ,c\}
,a\}+\{\{ c,a\} ,b\}=0
\end{array}
\end{equation}

\noindent and

\begin{equation}\label{eq25}
\begin{array}{rcl}
&\{ d,\{ c,a\}\}+\{ c,\{ a,d\}\}+[ a,\{ d , c\} ]=0,\cr &\{ d,\{ c
,b\}\}+\{ c,\{ b ,d\}\}+[b,\{ d,c\} ]=0,\cr &\{ a,\{ b,d\}\}+\{
b,\{ d,a\}\}+\{ d ,[a,b]\}=0,\cr &\{ a,\{ b ,c\}\}+\{ b,\{ c
,a\}\}+\{ c,[a,b]\}=0,
\end{array}
\end{equation}

\noindent are satisfied. These are the analogues in $\A$ of
expressions (\ref{eq10}) and (\ref{eq11}) for the $\de$-J-L
algebra $\jl$ in \cite{okam}.

\end{itemize}

In view of the novel and quite idiosyncratic normally ordered $\Z_{2}$-graded
associative multiplication structure $\circ$ of $\A$ (rules 0--2),
we must specify to the reader who wishes to verify patiently that
the graded Jacobi relations (\ref{eq24}) and (\ref{eq25}) hold how
to actually contract them. To this end, we define:

\noindent $\bullet$ {\bf Definition 2:} The contraction of a super-Jacobi relation is
said to be performed `{\em ($f$)rom ($i$)nside ($t$)o
($o$)utside}'\footnote{Write $fito$.} when the inner
$<.,.>$-brackets are opened and contracted first, and then the
outer ones. Analogously, the contraction of a super-Jacobi
relation is said to be $foti$ ({\it i.e.}, `{\em ($f$)rom
($o$)utside ($t$)o ($i$)nside}') when the outer brackets are
opened first, then the inner ones, and then the resulting
superpositions of words of length $3$ are totally contracted
according to rules 0--2.

\noindent $\bullet$ \underline{\bf Scholium:} The conscientious reader can check, by using
(\ref{eq13}), that {\em the super-Jacobi relations} (\ref{eq24}) and
(\ref{eq25}) {\em are satisfied by the $fito$ mode of contraction, but
not by the $foti$ one}. 

For instance, also to give an analytical
example of the two kinds of contraction, we evaluate the third
expression in (\ref{eq25}) by both $fito$ and $foti$ means

\begin{equation}\label{eq26}
\begin{array}{c}
\underline{fito}:~~\{ a,\{ b,d\}\}+\{ b,\{ d,a\}\}+\{ d
,[a,b]\}=\{ a, (bd)+(db)\}+\cr\{ b, (da)+(ad)\}+\{
d,(ab)-(ba)\}\stackrel{(\ref{eq13})}{=}\{ b, d-c\}=\{ b,d\}-\{
b,c\}=0\cr \underline{foti}:~~\{ a,\{ b,d\}\}+\{ b,\{ d,a\}\}+\{ d
,[a,b]\}=a\{ b,d\}+\{ b,d\} a +b\{ d,a\}+\cr\{ d,a\}b+
d[a,b]+[a,b]d= abd+adb+bda+dba+\cr
bda+bad+dab+adb+dab-dba+abd-bad= 2(ab)d+2(da)b+\cr 2(ad)b+2bda=
2(-db+da-cb-dba)=2(c+d-d+c)=2c\not=0.
\end{array}
\end{equation}

\noindent $\bullet$ \underline{\bf Result:} This indicates that, by virtue of the ordered
$\Z_{2}$-graded associative product structure of $\A$,

\begin{quotation}

\noindent{\em $\A$ is a Lie superalgebra-like structure with
respect to the $fito$, but not the $foti$, mode of contraction of
its graded Jacobi relations}.

\end{quotation}

\noindent This is another peculiar feature of $\A$---an immediate
consequence of its ordered $\Z_{2}$-graded associative
multiplication idiosyncracy.\footnote{In the next two sections we
will discuss in more detail these `multiplication oddities' of
$\A$.}

\section{Comparing $\jl$ with $\A$}\label{sec4}

We can now compare $\A$ with the abstract $\de$-J-L algebra $\jl$
defined in \cite{okam}. Below, we itemise this comparison:

\begin{itemize}

\item (i) As vector spaces, both $\jl$ and $\A$ are finite dimensional and
$\Z_{2}$-graded [(\ref{eq4}), (\ref{eq12})].

\item (ii) With respect to multiplication, while $\jl$ is
$\de$-associative ({\it i.e.}, associative $\slie$ for $\de=1$ or
antiassociative $\jor$ for $\de=-1$), $\A$ is multiplicatively {\sl NTO}-ordered and 
$\Z_{2}$-graded associative---a trait somewhat `in between' pure
associativity and pure antiassociativity [(\ref{eq6}),
(\ref{eq17}, \ref{eq18}, \ref{eq19})]; hence, 

\vskip 0.1in

\centerline{\underline{\em $\A$ is coined a hybrid $\de$-Jordan-Lie Superalgebra.}}

\vskip 0.1in

\item (iii) With respect to the $\Z_{2}$-graded commutation relations $<.,.>$, $\A$
combines characteristics of both Lie superalgebras
$\slie=\jl|_{\de=1}$ and J-L algebras $\jor=\jl|_{\de=-1}$. In
particular, as [(\ref{eq7}, \ref{eq9}), (\ref{eq21})] depict:

\begin{quotation}

\noindent ({\bf a}) $\A$ is like $\slie$ with respect to the
`homogeneous' $<.,.>$-relations obeyed by even and odd
elements.\footnote{That is to say, even elements obey antisymmetric
commutation relations, while odd elements obey symmetric
anticommutation relations. As noted earlier, this is a concise algebraic statement
of the celebrated {\em spin-statistics connection} \cite{pauli, www, freund}.}

\end{quotation}

\noindent while:

\begin{quotation}

\noindent ({\bf b}) $\A$ is like $\jor$ with respect to the
`inhomogeneous', `mixed spin-statistics' commutation relations between bosons and
fermions.\footnote{That is, the commutation relation between an
even and an odd element of $\A$, like in $\jor$, is symmetric
({\it i.e.}, the anticommutator bracket $<.,.>\equiv\{ .,.\}$).}

\noindent This is another novel feature of $\A$ as it reverses the usual $\Z_{2}$-graded superalgebraic spin-statistics connection rule, which maintains that the uncertainty relations between bosons (even generators) and fermions (odd generators) should be antisymmetric ({\it i.e.}, $<e,o>=[e,o]=eo-oe$) \cite{freund}. In our $\A$, even and odd generators obey symmetric anticommutator `uncertainty' relations: 

$$<e,o>=\{ e,o\}=eo+oe$$

\end{quotation}

\noindent moreover:

\begin{quotation}

\noindent ({\bf c}) The $\Z_{2}$-graded $<.,.>$-relations `close'
in $\A$ in exactly the same way that they close in
$\jl$\footnote{That is, in both $\jl$ and $\A$ the homogeneous
$<.,.>$-relations close in their even subspaces, while the
inhomogeneous ones in their odd subspaces.} [(\ref{eq8}),
(\ref{eq22})].

\end{quotation}

\item (iv) The generators of $\A$, unlike those in $\jl$, obey `externally
ungraded' Jacobi relations.\footnote{That is, the three external
factors $(-1)^{xz}$, $(-1)^{yx}$ and $(-1)^{zy}$ present in the
Jacobi expressions (\ref{eq10}) and (\ref{eq11}) for $\jl$ are
simply missing in the corresponding ones, (\ref{eq24}) and
(\ref{eq25}), for $\A$.} 

\noindent {\em In this formal respect, $\A$ is like an
ungraded Lie algebra $L$}.

\item (v) We return a bit to the comparison of the multiplication structure of the
two algebras (ii), now also in connection with the Jacobi
relations in (iv) above, and note that for the (anti)associative
$\de$-J-L superalgebras it is immaterial whether one evaluates
their super-Jacobi relations (\ref{eq10}) and (\ref{eq11}) $fito$
or $foti$, because they are `multiplicatively unordered'
structures.\footnote{That is, it does not matter in what order one
contracts pairs of generators in words of length greater than $2$
in $\jl$.}

On the other hand, as we saw in (\ref{eq26}) for
example, exactly because of the ordered $\Z_{2}$-graded
associative multiplication structure of $\A$, $fito$-contracted
Jacobis are satisfied in $\A$, but $foti$ ones are not, therefore
it crucially depends on the ordered multiplication structure
$\circ$ whether $\A$ is a Lie-like algebra ($fito$) or not
($foti$). Such a dependence is absent from the multiplicative
unordered $\slie$ and $\jor$ algebras.\footnote{The `multiplicative
unorderliness' (symmetry) of both $\jl|_{\de=1}=\slie$ and
$\jl|_{\de=-1}=\jor$ is encoded in the (anti)associativity
relation (\ref{eq6}) imposed on their products, since on the one
hand associativity simply means that the left-to-right contraction
of a $3$-letter word is the same as the right-to-left one, while
on the other, antiassociativity means essentially the same thing
under the proviso that one compensates with a minus sign for one
order of contraction relative to the other. Both associativity and
antiassociativity however, unlike the $\Z_{2}$-graded
associativity in $\A$ (\ref{eq19}), do not depend on the grade of
the letters involved in the binary contractions within words of
length greater than or equal to $3$. By contrast, the `multiplicative orderliness' or `directedness' 
(asymmetry) of our $\A$ comes hand in hand with the $\Z_{2}$-graded associativity of its binary 
product and the {\sl NTO}-ed mode of contraction of its three or more lettered words.}

\item (vi) Also in connection with (v) above, we note in view of
the lemma and the two corollaries concluding section \ref{sec2}
that:

\noindent ($\alpha$) Because $\A$ is not purely antiassociative,
words of length greater than or equal to $4$ in it do not vanish
identically as they do in $\jor$ for instance.\footnote{See lemma
in section \ref{sec2}.}

\noindent ($\beta$) Like the antiassociative $\jor$, $\A$ has no
idempotents and no two-sided identity element. However, as we saw in the
previous section, $\A$ has a right-identity, namely,
$a$.\footnote{See corollary 1 in section \ref{sec2}.}

\noindent ($\gamma$) As a corollary of $(\alpha$) above, and unlike
$\jor$, $\A$ is not nilpotent of length at most $4$.

\item (vii) Finally, in connection with (iii) and (iv) above, we note that
our choice of the symmetric anticommutator relation (as in $\jor$)
instead of the antisymmetric commutator relation (as in $\slie$)
for the inhomogeneous $<.,.>$-relations in $\A$ can be justified
as follows: had we assumed $[e,o]$ instead of $\{ e,o\}$, the
$fito$ contraction of the first super-Jacobi expression in
(\ref{eq24}) would yield

\[
\begin{array}{c}
[\{ d ,c\} ,a]+\{ [c,a],d\}+\{ [a,d],c\}=\{ c+d, d\}+\{ -c-d,
c\}=\cr \{ d,d\}-\{ c,c\}=-2a-2a=-4a\not=0,
\end{array}
\]

\noindent hence the graded Jacobi identities would not have been
obeyed by the generators of $\A$ and, as a result, the latter
could not qualify as an admissible Lie algebra \cite{lie1, lie, freund}.

\end{itemize}

We can distill the remarks above to the following important statement (result):

\begin{quotation}

\noindent $\bullet$ \underline{\bf Result:} The Lie admissibility of $\A$ ({\it i.e.}, the fact that $\A$ qualifies as an admissible (graded) Lie algebra obeying (graded) Jacobi identities) \cite{lie1, lie, freund} vitally depends on the fact that its algebraic binary product is normally ordered ({\sl NTO}-ed) and that its Jacobi anti-commutator relations are contracted in the $fito$, but not in the $foti$, order of contraction. Moreover, the super-Jacobi identities in $\A$ hold only if we assume symmetric (anticommutator) $\{ e,o\}$ and {\em not} anti-symmetric (commutator) $[e,o]$ commutation relations between generators of different grade.\footnote{Physically, these would correspond to quantum mechanical uncertainty relations between bosons and fermions \cite{freund}.}

\end{quotation}

\section{Closing remarks about $\A$}\label{sec5}

Our concluding remarks about $\A$ concentrate on the following
four issues:

\begin{itemize}

\item $\mathbf{(1)}$ We compare $\A$ against the other four possible Euclidean
division rings, namely, the reals ($\R$), the complexes ($\com$),
the quaternions ($\quat$) and the octonions ($\cayl$).

\item $\mathbf{(2)}$ As a particular case of $\mathbf{(1)}$, we remark about
the ordered $\Z_{2}$-graded associative $\A$ {\it versus} the
multiplicatively unordered, because purely associative, quaternions
$\quat$, and we briefly comment on the representation theory of
$\A$.

\item $\mathbf{(3)}$ We abstract $\A$ to a new type of Lie algebraic supervariety hitherto not encountered in the literature: a general
hybrid $\de$-Jordan-Lie superalgebra $\JL$, additionally possessing the novel multiplication structure of a normally ordered, $\Z_{2}$-graded associative, free linear semigroup.

\item $\mathbf{(4)}$ In the concluding Section 6, we discuss a possible physical
application and interpretation of $\A$ as originally anticipated
in \cite{rap1}.

\end{itemize}

\subsubsection{Comparing $\A$ to $\R$, $\com$, $\quat$ and $\cayl$}

\noindent $\mathbf{(1)}$ To make the aforesaid comparison, we first recall
how abstract algebraic structure gets lost upon climbing the
dimensional ladder from $\R$ to $\cayl$:

\begin{itemize}

\item Going from $\R$ of dimension $2^{0}=1$ to $\com$ of dimension $2^{1}=2$, one loses {\em
order}.\footnote{Although, one gains algebraic ploynomial solution completeness by solving equations such as $x^{2}+1=0$.}

\item Going from $\com$ of dimension $2^{1}=2$ to $\quat$ of dimension $2^{2}=4$, one loses {\em commutativity}.

\item Going from $\quat$ of dimension $2^{2}=4$ to $\cayl$ of dimension $2^{3}=8$, one loses {\em
associativity}.

\item And if one wished to extend the octonions to an algebra-like structure of dimension
$2^{4}=16$,\footnote{The formal procedure of extending $\com$ to $\quat$, $\quat$ to $\cayl$, and $\cayl$ to $\sed$, is known as the {\em Cayley-Dickson extension}. The algebra $\sed$ could be coined `{\em decahexanions}', but is more commonly known as the {\em sedenions} $\sed$ \cite{imaeda, sed}.} 
there would be no more abstract algebraic structure to be lost \cite{hur, kauff}.

\end{itemize}

We may subsume and organise the results of the comparison above into the following table

\begin{equation}\label{eq1}
\begin{tabular}{|c||c|c|c|c|}
\hline
Algebra &Order &L/R-Identity& Commutativity&Associativity \\ \hline\hline Reals ($\R$) &$\checkmark$ &$\checkmark$ &$\checkmark$ &$\checkmark$ \\
\hline Complexes ($\com$) &$\times$ &$\checkmark$ &$\checkmark$ &$\checkmark$ \\ \hline Quaternions  ($\quat$) &$\times$
&$\checkmark$ &$\times$ &$\checkmark$ \\ \hline Octonions ($\cayl$) &$\times$ &$\checkmark$ &$\times$ &$\times$ \\  \hline Alphabet ($\A$) &$\checkmark$ &$\times$ &$\times$ &$\times$ \\
\hline
\hline
\end{tabular}
\end{equation}

We can then summarise the comparison above by saying that $\A$ combines characteristics of all those four Euclidean
division rings $\R$, $\com$, $\quat$ and $\cayl$, in the following sense:

\begin{itemize}

\item (a) $\A$ is a vector space over $\R$.

\item (b) $\A$'s even subalgebra $\A^{0}$ is isomorphic to $\com$.

\item (c) $\A$ is a $4$-dimensional vector space like $\quat$, and its $3$-subspace spanned
by the mutually anticommuting $b$, $c$ and $d$ reminds one of the
subspace of real quaternions ({\it i.e.}, $\quat$ over $\R$) spanned
by the three imaginary ({\it i.e.}, $\sqrt{-1}$) quaternion units
$i$, $j$ and $k$.\footnote{With the important difference that $c$
in $\A$ is a `real', not an imaginary, unit ({\it i.e.}, $c=+\sqrt{a}\not=\sqrt{-a}$).}
Also, by comparing the multiplication tables (\ref{eq13}) and
(\ref{eq15}) for $\A$ and $\quat$ respectively, one immediately
realises that the former is a sort of {\em multiplicative deformation} of the
latter.\footnote{With most notable `deformation features' of the
generators of $\A$ relative to those of $\quat$ being $c$'s
squaring to $a$ unlike $j$'s squaring to $-1$ mentioned in the
last footnote, and $a$'s role only as a right-identity unlike
$1$'s role in $\quat$ as a two-sided identity.
In fact, from the diagonals of their respective multiplication
tables (\ref{eq15}) and (\ref{eq13}), one could say that the unit
quaternions in $\mathcal{U}$ naturally support a metric of {\em
Lorentzian signature} $\mathrm{diag}(1,-1,-1,-1)$ (:absolute trace
2) \cite{lambek, trif, trif1}, while the units of $\A$ in $\mathcal{G}$
support a metric of {\em traceless Kleinian signature}
$\mathrm{diag}(1,-1,1,-1)$. See ensuing discussion on $\A$'s norm below, as well as the
correspondence (\ref{eq14}) in section \ref{sec3}.}

\item (d)  Like the algebra of octonions $\cayl$, $\A$ is not associative.\footnote{See further, more detailed remarks on Associativity in the next subsection.}

\item (e) Furthermore, the novel multiplicatively normally ordered ({\sl NTO}-ed), 
$\Z_{2}$-graded associative structure of $\A$ recalls a bit the
linearly ordered $\R$.

\end{itemize}

\subsubsection{Comments on possible `matrix' representations of $\A$}

\noindent $\mathbf{(2)}$ We stressed above the close similarities between
$\A$ and $\quat$. Now we would like to gain some more insight into
the novel non-associativity of $\A$ by comparing it with the
associative quaternions. As a bonus from such a comparison, we
will also comment briefly on a possible representation of $\A$.

So, we may recall from \cite{lambek} the real 4-dimensional left
($L$) and right ($R$) matrix `self-representations'\footnote{The
epithet `self' refers to the representation of $\quat$ (by real
matrices) induced by the quaternions' own algebraic product.} of
quaternions over $\R$

\begin{equation}\label{eq27}
{\rm Left:}~ab=c\mapto L(a)[b]=[c]~~{\rm and}~~{\rm
Right:}~bc=d\mapto R(c)[b]=[d],
\end{equation}

\noindent where $[b]$ is a column vector in $\R^{4}$,\footnote{That
is, in the expansion of the real quaternion $b$ in the standard
unit quaternion basis $\mathcal{U}$:
$b=b_{0}1+b_{1}i+b_{2}j+b_{3}k$, the entries of the $4$-vector
$[b]$ are the real numbers $b_{\mu}$.} while both $L(a)$ and
$R(c)$ are $(4\times 4)$-real matrices.\footnote{It is easy to check
that the maps $L$ and $R$ are homomorphisms of $\quat$ ({\it i.e.},
representations of $\quat$).} The crucial point is that, because
$\quat$ is associative,

\begin{equation}\label{eq28}
(ab)c=a(bc)\Rightarrow R(c)L(a)[b]=L(a)R(c)[b]\Leftrightarrow
[L(\quat),R(\quat)]=0,
\end{equation}

\noindent and similarly, for a purely antiassociative algebra like
$\jor$ before, it follows that

\begin{equation}\label{eq29}
\{ L(\jor), R(\jor)\}=0.
\end{equation}

\noindent We may summarise (\ref{eq28}) and (\ref{eq29}) to the
following:

\begin{quotation}

\noindent The left and right self-representations of an {\em
associative} algebra {\em commute}, while those of an {\em
antiassociative} algebra {\em anticommute}.

\end{quotation}

\noindent It follows that the self-representations of $\A$, which
is neither purely associative nor purely associative (but somewhat
in between the two), will neither commute nor anticommute with
each other. As a matter of fact, since $\A$ is multiplicatively
ordered $frtl$, only its left self-representation would be
relevant (if it actually existed\footnote{This author has not been
able to construct yet a matrix representation of $\A$ based on its
ordered $\Z_{2}$-graded associative product. {\it In toto}, since our $\A$ is nonassociative, 
one would expect it {\em not} to have a standard linear (:matrix) representation in $M_{4}(\R)$, as all (real) matrix algebras are associative under 
matrix multiplication. Of course, like with
all the usual Lie algebraic varieties and supervarieties, we could
alternatively look directly into a possible representation of the
non-associative (under the Lie bracket $<.,.>$ now) $\A$ by a
(possibly graded) Lie algebra $End(V)$ of endomorphisms of a
suitable (possibly graded) vector space $V$. However, this alternative has not been
seriously pursued or explored yet.}).

\subsubsection{Abstracting and generalising $\A$}

\noindent $\mathbf{(3)}$ The abstraction of $\A$ to a general hybrid
$\de$-J-L algebra $\JL$ is straightforward:

\begin{quotation}

\noindent A finite dimensional $\Z_{2}$-graded vector space $\JL$
over a field $K$ of characteristic not $2$, together with a normally
ordered $\Z_{2}$-graded associative free algebraic product between its generators and a bilinear $\Z_{2}$-graded Lie-like bracket $<.,.>$ satisfying
(\ref{eq21})--(\ref{eq25}), is called an {\em abstract hybrid
$\de$-Jordan-Lie superalgebra}.

\end{quotation}

In fact, since the binary product $\circ$ in $\A$ is normally ordered and `directed' {(:from right to left) in the {\sl NTO}-ed sense, it resembles a semigroup product; albeit, {\em a free, nonassociative} (:$\Z_{2}$-graded associative) {\em linear semigroup over} $\R$, subject to the aforementioned free relations (or product concatenation) and contraction rules.

\hskip 0.1in

Thus, all in all:

\begin{quotation}

\noindent {\em Our alphabetic $\A$ is an instance of an abstract hybrid
$\de$-Jordan-Lie superalgebra and a free, multiplicatively normally ordered, $\Z_{2}$-graded associative, linear semigroup}.

\end{quotation}

\noindent Before we give our last remarks on a possible physical application of $\A$ in the concluding section, we turn our attention to matters of {\em associativity} of an algebraic binary product so as to shed more light on the novel and quite peculiar binary multiplication structure $\circ$ of $\A$.

\subsection{Varia on Associativity}

Now that we have seen that extending the quaternions $\quat$ to the octonions $\cayl$ results in losing associativity of the algebraic product, we recall that if one followed a 
general Cayley-Dickson complexification-type of formal procedure for further extending the algebra of octonions $\cayl$ to the $2^{4}=16$-dimensional algebra of the {\em sedenions} $\sed$ \cite{imaeda, sed}, and, moreover, doubled the latter to the $2^{5}=32$-dimensional algebra $\trig$ of the so-called {\em trigintaduonions} \cite{trig1, trig}, we would witness the following progressive weakening of associativity.

\subsubsection{Alternativity}

When one extends $\quat$ to $\cayl$ \cite{conway, lounesto, springer}, one loses associativity; however, the octonion binary product still obeys a weaker form of associativity coined {\em alternativity}. That is to say,

 \begin{equation}\label{eq30}
\forall x,y\in\cayl :\,\,\left\lbrace\begin{array}{rcl}  (xx)y=x(xy),& \mbox{Left-Alternativity},\cr y(xx)=(yx)x, & \mbox{Right-Alternativity}.
\end{array} \right.
\end{equation}

\hskip 0.1in

\noindent $\bullet$ \underline{\bf Result:} Our alphabet algebra $\A$ is not alternative.  Here is a sample calculation, always following the multiplication table of the generators of $\A$ in (\ref{eq13}), showing the violation of Left-Alternativity in $\A$:

\begin{equation}\label{eq31}
d(dc)=db=-c,\,\,\, (dd)c=-ac=d\,\Rightarrow\, d(dc)\not=(dd)c
\end{equation}

\subsubsection{Power Associativity}

If one further extends the octonions $\cayl$ to the sedenions $\sed$ \cite{imaeda, sed}, one loses even alternativity; however, the sedenion product still obeys a weaker form of associativity coined {\em power associativity}. That is to say,

 \begin{equation}\label{eq32}
\forall x\in\sed :\,\, (xx)x=x(xx)
\end{equation}

\noindent and for powers of $x$ greater than $3$.\footnote{For instance, for $x^{4}$, for all $x$ in a power associative algebra, we would observe: $(xx)xx=x(xx)x=xx(xx)$.}

\hskip 0.1in

\noindent $\bullet$ \underline{\bf Result:} Our alphabet algebra $\A$ is not power associative.  Again, here is a sample calculation, always following the multiplication table of the generators of $\A$ in (\ref{eq13}), showing the violation of Power Associativity in $\A$:

\begin{equation}\label{eq33}
d(dd)=-da=-d,\,\,\, (dd)d=-ad=c\,\Rightarrow\, d(dd)\not= (dd)d
\end{equation}

\subsubsection{Flexibility}

Finally, if one wished to further extend the sedenions $\sed$ to the algebra of {\em trigintaduonions} $\trig$ \cite{trig1, trig}, one should further relax power associativity to a still weaker form of associativity coined {\em flexibility}. That is to say,

 \begin{equation}\label{eq34}
\forall x, y\in\trig :\,\, (xy)x=x(yx)
\end{equation}

\hskip 0.1in

\noindent $\bullet$ \underline{\bf Result:} Our alphabet algebra $\A$ is flexible.  One need only check, always using (\ref{eq13}), that the following twelve equations hold between the four generators of $\A$:

\begin{equation}\label{eq35}
\begin{array}{rcl}
&a(ba)=(ab)a\cr 
&b(ab)=(ba)b\cr
&a(ca)=(ac)a\cr 
&c(ac)=(ca)c\cr 
&a(da)=(ad)a\cr
&d(ad)=(da)d\cr
&b(cb)=(bc)b\cr
&c(bc)=(cb)c\cr
&b(db)=(bd)b\cr
&d(bd)=(db)d\cr
&c(dc)=(cd)c\cr
&d(cd)=(dc)d\cr
\end{array}
\end{equation}

\hskip 0.1in

\noindent $\bullet$ We may organise the results of the comparisons above into the following table:

$$
\begin{tabular}{|c||c|c|c|c|}
\hline
Algebra &Associativity &Alternativity& Power Associativity&Flexibility\\ \hline\hline Quaternions $\quat$ &$\checkmark$ &$\checkmark$ &$\checkmark$ &$\checkmark$ \\
\hline Octonions $\cayl$ &$\times$ &$\checkmark$ &$\checkmark$ &$\checkmark$ \\ \hline Sedenions $\sed$ &$\times$
&$\times$ &$\checkmark$ &$\checkmark$ \\ \hline  Trigintaduonions $\trig$&$\times$ &$\times$ &$\times$ &$\checkmark$ \\ \hline
Alphabet $\A$&$\times$ &$\times$ &$\times$ &$\checkmark$ \\
\hline
\end{tabular}
$$

\hskip 0.1in

\noindent Thus, we observe that {\em in matters of algebraic product associativity, our $4$-dimensional alphabet algebra $\A$ is more like the $32$-dimensional algebra of trigintaduonions} $\trig$.

\subsection{Miscellaneous Structural-Algebraic Matters on Identity, Involution, Norm and Lie Admissibility}

\subsubsection{Matters of Identity, Involution and Norm}

All four `generalised number' algebras, regarded as algebras over the field of reals $\R$: the real numbers $\R$ themselves, the complexes $\com$, the quaternions $\quat$ and the octonions $\cayl$ are commonly knows {\em Euclidean division rings}, because:

\begin{itemize}

\item They all have a multiplicative $2$-sided identity: namely, the number $1$.

\item $\com$, $\quat$ and $\cayl$ all have a unary operation, called {\em conjugation} $\star :\, A\rightarrow A$, which, to every element $x$ in the corresponding algebra $A$, assigns its conjugate $x^{\star}$ in $A$ (which can be readily seen to commute with $x$ itself).\footnote{For $\com$, $\star$ is simply the complex conjugation map, while for the reals $\R$, $\star$ simply reduces to the identity map as every real number is self-conjugate.}

\item For the non-commutative $\quat$ and $\cayl$, conjugation acts as an {\em involution}, in the sense that it reverses the order of the algebraic product of the corresponding elements: $(x_{1}x_{2})^{\star}=x_{2}^{\star}x_{1}^{\star}$.\footnote{$\com$ is a commutative algebra, hence conjugation does not affect the order of multiplication.}

\item Having defined conjugation, the {\em norm} of the elements in $\com$, $\quat$ and $\cayl$ is defined via the act of conjugation to be: $\mathcal{N}(x)\equiv\lVert x\rVert =\sqrt{xx^{\star}}=\sqrt{x^{\star}x}$.

\item Furthermore, the norm is seen to be {\em real Euclidean}, of positive definite signature: $\mathcal{N}(\cdot )=\lVert \cdot\rVert : A\rightarrow \R_{+},\,\,\forall x\in A:\, \lVert x\rVert\geq 0$.\footnote{In $\R$, the norm is simply the absolute value of a real number.} This is why all four division rings $\R$, $\com$, $\quat$ and $\cayl$ are called {\em Euclidean}.

\item Finally, all four Euclidean division rings above are {\em composition algebras} $\mathcal{A}$, in the following defining sense:

\end{itemize}

\begin{equation}\label{eq36}
\mathcal{N}(xy)=\mathcal{N}(x)\mathcal{N}(y)\, (\forall x,y\in\mathcal{A})
\end{equation}

\noindent where $\mathcal{N}(x)\equiv\lVert x\rVert$ is {\em the norm of $x$}, as defined above.

\hskip 0.1in

$\bullet$ {\bf By contrast, $\A$...}: 

\begin{itemize}

\item $\A$ does not have a multiplicative $2$-sided identity: the generator $a$ serves only as a right-identity (\ref{eq13}).

\item $\A$ does not have a conjugation operation that reverses the order of multiplication of its generators. On the contrary, $\A$ is a {\em $\Z_{2}$-graded, lexicographically/normally ordered multiplicative structure} (semigroup).

\item In $\A$, we can define a `pseudo-norm' type of map, not via a conjugation/involution unary operation as in the other four Euclidean division rings, but simply by squaring each generator and taking the (real number) coefficient of its square: $\lVert x\rVert =\mathrm{Coeff}_{\R}(x^{2})$.\footnote{This recalls a bit how an abstract {\em Clifford Algebra} $Cl(V^{n},g)$ is generally defined \cite{hestenes}, as {\em an $n$-dimensional vector space $V^{n}$ (over a field of characteristic not $2$), equipped with a symmetric bilinear form (inner product) $g(x,y)$, that is freely generating a (real or complex) associative algebra $Cl(V^{n},\cdot ,g)$ subject to the anticommutation relation}: $\forall x,y\in Cl:\,\{ x,y\}=x\cdot y + y\cdot x=2g(x,y)\mathbf{1}$, where $x\cdot y=xy$ is simply the unital and associative binary algebraic product in $Cl$. When $x=y$, it follows that: $\{ x,x\}=x\cdot x+x\cdot x= xx+xx=2x^{2}=2g(x,x)\mathbf{1}\Rightarrow \lVert x\rVert^{2}:=g(x,x)=\mathrm{Coeff}_{\R}(x^{2})$.}

\item With this pseudo-norm type of map, from the diagonal of the multiplication table (\ref{eq13}), the reader immediately notices the the norms of the four generators of $\A$ are: $\lVert a\rVert =+1$, $\lVert b\rVert =-1$, $\lVert c\rVert =+1$ and $\lVert d\rVert =-1$, respectively.

\item Thus, the pseudo-metric $\eta$ that is naturally associated with the pseudo-norm on the natural standard basis of generators of $\A$ above is an indefinite, Kleinian-type of metric, of $0$-signature: $\eta_{\mu\nu}=(1,-1,1,-1)$; $\mathrm{tr}(\eta_{\mu\nu})=0$ (traceless).

\item $\A$ {\em is manifestly not a composition algebra} ({\it i.e.}, $\mathcal{N}(xy)\not=\mathcal{N}(x)\mathcal{N}(y)$), as one can straightforwardly verify.

\item Finally, and {\it en passant}, we note that, with the pseudo-norm defined above, we can readily find four {\em null} (:nilpotent) vectors in $\A$, namely: $n_{1,2}=d\pm c$ and $n_{3,4}=c\pm b$, as one can readily verify that $n_{\mu}^{2}=0$. For example, for $\mu=1$, $\lVert n_{1}\rVert =\mathrm{Coeff}_{\R}(n_{1}^{2})= \mathrm{Coeff}_{\R}[(d+c)^{2}]=\mathrm{Coeff}_{\R}[(d+c)(d+c)]=\mathrm{Coeff}_{\R}[d^{2}+dc+cd+c^{2}]=\mathrm{Coeff}_{\R}[-a+b-b+a]=0$.

\end{itemize}

\subsubsection{Lie Admissibility and Brief Comparison to Okubo Algebras}

We witnessed above how our alphabetic algebra $\A$ combines structural characteristics from, and in a way extends, all the four Euclidean devision rings $\R$, $\com$, $\quat$ and $\cayl$, and their extensions to $\sed$ and $\trig$, plus we have seen that it admits a $\Z_{2}$-graded Lie bracket type of bilinear product with respect to which it qualifies as a {\em Lie admissible algebra} \cite{lie}.

We then saw that, with respect to the said $\Z_{2}$-graded Lie bracket, certain $fito$-ordered $\Z_{2}$-graded Jacobi identities are satisfied, which make $\A$ qualify as a $\de$-Jordan-Lie Superalgebra in the sense of Okubo and Kamiya \cite{okam}.

From a more general vantage, $\A$ may be regarded as a multiplicatively ordered and $\Z_{2}$-graded associative version of {\em Okubo algebras}, which are non-Euclidean (:pseudo-metric) generalisations of the quaternions $\quat$ and the octonions $\cayl$ \cite{okubo1, okubo2, okubo3}. To make further analogies with our $\A$ in view of our associativity remarks earlier, {\em Okubo algebras are non-associative composition algebras, flexible algebras, Lie admissible algebras, power associative, yet non-alternative algebras, and, like our $\A$, they do not have a $2$-sided identity element}.

\section{Heuristic Smatterings on a Possible Application to Theoretical Physics}

We conclude the present paper by allowing ourselves
some leeway and latitude so as to discuss briefly a possible physical
application and concomitant interpretation of $\A$.

\vskip 0.07in

The alphabetic algebra $\A$ was originally conceived in this author's Ph.D. thesis \cite{rap1}, but not in the
rather sophisticated $\de$-J-L superalgebra guise presented above. The basic
intuition in \cite{rap1} was to give a simple `{\em generative
grammar}'-like theoretical scenario for the creation of spacetime
from a finite number of quanta (generators) which were supposed to
inhabit the quantum spacetime substratum commonly known as the
vacuum \cite{df91}. Thus, it was envisaged that a spacetime-like
structure could arise from the algebraic combinations of a finite
number of quanta, as it were, a combinatory-algebraic process
modelling the {\it aufbau} of spacetime from quantum spacetime
numbers filling the vacuum.\footnote{Thus, $\A$ could be coined
`quantum spacetime arithmetic' and the imagined process of
building spacetime from such abstract numbers is akin, at least in
spirit, to how relativistic spacetime was assembled from abstract
digits and a suitable code or `algorithm' for them in
\cite{df69}.} Furthermore, by the very alphabetic character of
$\A$ and its alphabetically ordered algebraic structure, this
syntactic {\em lexicographic} process representing the building of
spacetime was envisaged to encode the germs of the primordial
`{\em quantum arrow of time}' in the sense that a primitive
`temporal directedness' is already built into the algebraic
structure of those quantum spacetime numbers---a basic order or
`{\it taxis}' inherent in the very rules for the algebraic
combinations of the generators of $\A$, as we saw before. 

In view
of the intimate structural similarities between $\A$ and the
quaternion division algebra $\quat$ mentioned above, and since the
latter are so closely tied to the structure of relativistic
spacetime and the best unification between quantum mechanics and
(special) relativity that has been achieved so far, namely, the Dirac
equation \cite{lambek}\footnote{For example, in \cite{lambek, lambek2}
Minkowski vectors are represented by hermitian biquaternions,
Lorentz transformations by unimodular complex quaternions
(essentially, the biquaternion analogues of the elements of
$SL(2,\com)$---the double covering of the Lorentz group), the
$3$-generators $\sigma_{i}$ of the Pauli spin Lie algebra $su(2)$
are just the three mutually anticommuting `imaginary' quaternions
multiplied by the complex number $i$ in front ($i^{2}=-1$), and,
most importantly, the Dirac equation can be derived very simply
and entirely algebraically from $\quat$ over $\com$ ({\it i.e.},
from biquaternions). Also, as noted in footnote 40, the Lorentzian
signature (and even the $4$-dimensionality!) of Minkowski spacetime is
effectively encoded in (the diagonal of) the multiplication table
(\ref{eq15}) of the unit quaternions in $\mathcal{U}$
\cite{trif, trif1}.}, we can imagine that $\A$ could be somehow used in
the future to represent algebraically a `time-directed' sort of
Minkowski spacetime and, therefore, a time-asymmetric version of the Dirac
equation that would appear to be supported rather naturally by the
former. 

In this line of thought,  in a forthcoming paper \cite{rap2} we entertain the possibility of arriving at an inherently and genuinely {\em Time-Asymmetric Dirac Equation} by entirely algebraic means. This project comes to answer a 70 years' old conundrum that Lambek faced when he first derived the Dirac Equation using quaternions, as posited in \cite{lambek, lambek2}: 

\begin{quotation}

\noindent {\em To use real quaternions over the field $\R$ of real numbers, or complex quaternions over the field $\com$ of complexes?} 

\end{quotation}

In \cite{lambek, lambek2}, Jim Lambek very tellingly recalls how he told Paul Dirac back in the 50s that he could derive his famous equation for the electron's dynamics using (admittedly, complex) quaternions (over $\com$)---commonly referred to as {\em biquaternions}---for which Dirac appeared not to have been greatly impressed.\footnote{A reaction that, as Lambek recalls in \cite{lambek}, unfortunately discouraged him from further pursuing his Mathematical/Theoretical Physics interests, and rather focus his attention solely on Mathematics.} Four decades later, upon writing \cite{lambek}, Jim came to `regret' not to have challenged back Dirac, by saying: ``...{\em alright, but can \underline{you} derive your equation using real quaternions over the field $\R$ of real numbers?}''. Of course, we do now have real spinor representations of the Dirac $\Gamma$-matrices, called {\em Majorana representations} \cite{nlab}, but that is {\em not} what Lambek meant to challenge Dirac about. Lambek simply wanted to find out whether Dirac could derive his equation from real quaternions (over $\R$) alone, not whether there are real representations of the spinorial wave functions involved in it. Our hybrid $\de$-Jordan-Lie superalgebra $\A$ may be able to address this `deficiency' and arrive at a genuinely real (:$\A$ is an {\em algebra} over $\R$) Dirac-type of equation which, {\it a fortiori}, is {\em inherently time-asymmetric}.}

However, the quest in this direction is far from its
completion.

\vskip 0.05in

We would like to close the present paper in the spirit of the last
paragraph with a suitable quote from the end of \cite{kauff} that, in a sense, vindicates our perspective on $\A$ as a multiplicatively ordered and directed `{\em free generative algebra}':

\vskip 0.2in

\centerline{``In the beginning was the word.}

\centerline{The word became self-referential/periodic.}

\centerline{In the sorting of its lexicographic orders,}

\centerline{The word became topology, geometry and}

\centerline{The dynamics of forms;}

\centerline{Thus were chaos and order}

\centerline{Brought forth together}

\centerline{From the void.''}

\centerline{(from {\bf CODA})}

\section*{Acknowledgments}

Some early `prophetic' remarks by Jim Lambek about $\quat$ in \cite{lambek} almost three decades ago
and in subsequent private correspondence before the year 2004, helped this author
clarify by analogy and juxtaposition some crucial structural-algebraic features of $\A$.

\end{document}